\documentclass[aps,twocolumn,showpacs,preprintnumbers,nofootinbib,prd,superscriptaddress,groupedaddress,10pt]{revtex4-1}
\usepackage[utf8]{inputenc}

\makeatletter
\def\l@subsubsection#1#2{}
\def\l@subsubsubsection#1#2{}
\makeatother

\usepackage{amsmath}
\usepackage{graphicx,amssymb,amsmath,amsthm,amsfonts,epsfig,epsf}
\usepackage[usenames,dvipsnames,svgnames,table]{xcolor}
\usepackage{epstopdf}
\definecolor{darkred}{rgb}{0.5,0,0}
\usepackage{aas_macros}
\usepackage{bm}
\usepackage{dcolumn}
\usepackage[utf8]{inputenc}
\usepackage{latexsym}
\usepackage{rotating}
\usepackage{longtable}

\usepackage{enumerate}
\usepackage{tensor,multirow}
\usepackage{mathtools}
\usepackage{url}
\usepackage[linktocpage,hidelinks]{hyperref}

\newcommand{\mndd}{_{\mu\nu}}

\newcommand{\ijdd}{_{ij}}
\newcommand{\ijuu}{^{ij}}

\newcommand{\til}{~}

\def\nn{\nonumber}

\def\be{\begin{equation}}

\newcommand{\nocontentsline}[3]{}
\newcommand{\tocless}[2]{\bgroup\let\addcontentsline=\nocontentsline#1{#2}\egroup}
\def\ba{\begin{align}}
\def\ea{\end{align}}

\newcommand{\warn}[1]{{\textcolor{red}{\sf{[IN PROGRESS]}} }}

\begin{document}

\title{
Applications of the close-limit approximation: horizonless compact objects and scalar fields\\
}

\author{
Lorenzo Annulli$^{1}$,
Vitor Cardoso$^{1}$,
Leonardo Gualtieri$^{2}$
}

\affiliation{${^1}$ Centro de Astrof\'{\i}sica e Gravita\c c\~ao - CENTRA, Departamento de F\'{\i}sica, Instituto Superior T\'ecnico - IST, Universidade de Lisboa - UL, Avenida Rovisco Pais 1, 1049-001 Lisboa, Portugal}
\affiliation{${^2}$
Dipartimento di Fisica, ``Sapienza'' Universit\`a di Roma and\\ Sezione INFN Roma1, Piazzale Aldo Moro 5, 00185 Roma, Italy}

\begin{abstract}
The ability to model the evolution of compact binaries from the inspiral to coalescence is central to gravitational wave
astronomy. Current waveform catalogues are built from vacuum binary black hole models, by evolving Einstein equations
numerically and complementing them with knowledge from slow-motion expansions.
Much less is known about the coalescence process in the presence of matter, or in theories other than General
Relativity.  Here, we explore the Close Limit Approximation as a powerful tool to understand the coalescence process in
general setups.  In particular, we study the head-on collision of two equal-mass, compact but horizonless objects. Our
results show the appearance of ``echoes'' and indicate that a significant fraction of the merger energy goes into these
late-time repetitions.  We also apply the Close Limit Approximation to investigate the effect of colliding black holes
on surrounding scalar fields. Notably, our results indicate that observables obtained through perturbation theory may be
extended to a significant segment of the merger phase, where in principle only a numerical approach is appropriate.
\end{abstract}

\maketitle


\section{Introduction}\label{sec:Introduction}
A few years ago we witnessed the birth of a new science, gravitational wave (GW)
astronomy~\cite{TheLIGOScientific:2016src,Abbott:2016nmj,Abbott:2017vtc,Abbott:2017oio,Abbott:2020tfl}.  Observatories
like LIGO and Virgo are now able to see routinely the inspiral and coalescence of compact objects such as black holes
(BHs) and neutron stars.  Observations are so far consistent with all the predictions made by General Relativity (GR), even
in regimes probing strong-field and dynamical processes.  In the near future, similar observations will be pushed to exquisite precision with third-generation detectors and the space-based LISA mission.  With high quality data and low
instrumental noise, these observatories will have access to virtually all the visible
universe~\cite{Maggiore:2019uih,Audley:2017drz}.

In this new era of GW astronomy, new precision tests of GR and search for new physics also become
possible~\cite{Berti:2015itd,Barack:2018yly,Cardoso:2019rvt}.  Of special relevance is the scrutiny of GR predictions
regarding the gravitational interaction in the strong-field regime.  A fundamental result of vacuum GR is that isolated
BHs all belong to the same family of solutions -- the Kerr family~\cite{Kerr:1963ud} -- fully described by two
parameters alone, mass and angular momentum~\cite{Chrusciel:2012jk,Robinson:2004zz,Cardoso:2016ryw}.
Thus, testing the Kerr nature of BHs means, to some extent, testing GR.
In addition, foundational questions are associated with the presence of horizons, particularly issues concerning the
breakdown of determinism associated with Cauchy horizons or the fate of singularities of the classical
equations~\cite{Penrose:1964wq,Penrose:1969,Cardoso:2017soq,Cardoso:2017cqb,Cardoso:2019rvt}. While possible
pathological behavior is conjectured to be hidden behind horizons, questions remain concerning the effect of quantum
gravity on the near-horizon structure or even on horizons themselves: do horizons exist?  Are the objects we observe
really BHs, or are they extreme (and exotic) compact objects (ECOs) which mimic the BH behaviour?  Gravitational wave
astronomy can have an important role in this matter, by constraining the existence of ``echoes'' or assessing the tidal
properties of compact
objects~\cite{Cardoso:2016rao,Cardoso:2016oxy,Abedi:2016hgu,Nielsen:2018lkf,Abedi:2018pst,Lo:2018sep,Tsang:2018uie,Uchikata:2019frs,Abbott:2020jks,Wang:2020ayy,Maselli:2017cmm,Agullo:2020hxe,Cardoso:2019rvt}.
Precision GW astronomy can also inform us on the nature and distribution of dark matter. A nontrivial dark matter
environment changes the inspiral of a compact binary, via accretion or dynamical friction. If dark matter consists on
new fundamental light fields, then rotating BHs can become lighthouses of
GWs~\cite{Brito:2015oca,Hui:2016ltb,Bertone:2018krk,Baibhav:2019rsa,Annulli:2020lyc}.

The possibility of
extracting meaningful results and information from GW observations depends on our ability to model the coalescence of
compact objects. These processes are now well understood: the inspiral part is modeled using slow-motion
expansions~\cite{Blanchet:2013haa,Poisson_will_2014}, the merger through fully non-linear numerical simulations (the
so-called ``numerical relativity'' (NR))~\cite{Pretorius:2005gq,Campanelli:2005dd,Baker:2005vv,Buchman:2012dw}, and the
ringdown of the final object is modeled using spacetime perturbation
theory~\cite{Regge:1957td,Zerilli:1971wd,Kokkotas:1999bd,Ferrari:2007dd,Berti:2009kk}. Perturbation theory, extended to
include self-force effects~\cite{Barack:2009ux,Poisson:2011nh}, can also be used to model extreme mass ratio binaries. 

These models are built in the framework of GR, and rely on the hypothesis that all compact objects in the Universe (with
masses $\gtrsim(2.5-3)\,M_\odot$) are BHs. To test GR and the BH hypothesis beyond null or biased
tests~\cite{Yunes:2009ke}, one needs to develop models of the inspiral, merger and ringdown which do not assume GR and do
not assume that the compact objects are BHs. In particular, to test GR and the BH hypothesis in the strong-field regime
we need to extend the models of the merger phase. However, the extension of NR to modified gravity theories and to ECOs
is not an easy task, since they require a well-posed and well-behaved formulation of the time-evolution
problem~\cite{Delsate:2014hba,Papallo:2017qvl,Ripley:2019irj,Kovacs:2020ywu,Witek:2020uzz,East:2020hgw}.

A possible way to circumvent the above problem is to use an alternative approach, such as the close-limit
approximation
(CLAP)~\cite{Price:1994pm,Abrahams:1995wd,Nicasio:1998aj,Khanna:1999mh,Gleiser:1996yc,Gleiser:1998rw,Allen:1999rg,Sopuerta:2006wj,LeTiec:2009yf}.
In this approach, the slice of spacetime describing a late stage of the
merger is computed solving the appropriate constraints of the field equations, as is commonly done to find the initial
data for NR simulations. These initial data, if the two bodies are close enough, can be interpreted as a deformation of
the stationary spacetime describing the final BH. Therefore, the subsequent evolution of this geometry can
be studied using perturbation theory: the perturbation parameter, in this case, is the separation between the two BHs.
Thus, we do not need NR to evolve these initial data: it is sufficient to solve the perturbation equations (e.g. the
Zerilli equation) to obtain the GWs emission from the merger and ringdown stages. Additionally, a recent method called the ``Backwards One-Body (BOB) method''\til\cite{McWilliams:2018ztb} assumes as well the end state of a binary coalescence to compute waveforms emitted by coalescing  binaries; in this context, the dynamics of light rays around the final perturbed BH is the key to compute the full GW signal.

Since the CLAP showed a remarkable agreement with NR simulations~\cite{Anninos:1993zj,Anninos:1998wt},
its extension beyond GR and beyond the BH hypothesis could be a valuable tool to model the merger stages of compact
binary coalescence in an extended framework.

This article is the first step towards the extension of the CLAP to modified theories of gravity and to extreme compact
objects. In the first part of the paper we shall consider head-on collisions of ECOs, finding the gravitational waveform
emitted in this process. In this article we shall consider the simplest version of the CLAP, in which the initial data
are assumed to be static. This is of course a strong approximation, which will be relaxed in subsequent work. Thus, our
results should be interpreted only as an indication, showing the qualitative features of the emitted GW signal.  We also
remark that, since we consider the head-on collision of non-spinning objects, we do not capture the angular momentum
emission which would be produced in a more realistic setup.

%
In the second part we shall consider a test scalar field in a binary BH (BBH) spacetime (which can be interpreted either
as ultra-light fundamental fields, or as extra gravitational degrees of freedom in scalar-tensor gravity), finding an
indication on how the dynamical evolution of the background during the merger --~which we model using the CLAP~--
affects the quasi-normal mode (QNM) frequencies of the scalar field.

These are just the first steps of this project: in future works we shall implement non-head-on and non-static initial
data, to study the emission from the merger of two ECOs, and the emission of a BBH merger in modified theories of
gravity such as Einstein-dilaton Gauss-Bonnet gravity or dynamical Chern-Simons gravity.

In Sec.~\ref{sec:CLAP} we review the standard CLAP approach to describe the merger of BBHs in GR. In
Sec.~\ref{sec:BH_mimickers} we use the CLAP to describe the head-on collision of two ECOs. In Sec.~\ref{sec:CLAP_STT} we
use the CLAP to show the effect of a BBH merger on the QNMs of a test scalar field. In the following we use geometric
units $G=c=1$.  Greek indices refer to quantities defined on a four-dimensional ($4D$) spacetime manifold, Latin indices
to their three-dimensional spatial parts ($3D$).

\section{The close limit approximation}\label{sec:CLAP}
We shall here review the standard CLAP approach for BBH coalescences in GR.  This approach is based on the assumption
that in the final stages of the coalescence, when the two BHs are sufficiently close to each other, the spacetime is a
small deformation of a single BH spacetime, and thus can be studied using the techniques of spacetime perturbation
theory. To this aim, BBH initial data, originally developed in the context of NR, are recast as a perturbation of Kerr spacetime. In this article we shall consider, as a first step, the case of a head-on collision of non-rotating BHs, which
leads to a non-rotating BH. Thus, the BBH initial data are recast as a perturbation of a Schwarzschild BH.

\subsection{The $3+1$ decomposition}\label{sec:3+1}
We briefly recall the basic concepts of the $3+1$ decomposition in NR. We refer the interested reader to one of
the several excellent books and reviews on the
subject\,\cite{Gourgoulhon:2007ue,alcubierre2008introduction,baumgarte2010numerical,shibata2015numerical} for further
information and details.

In order to formulate GR (or any other gravitational theory) as a time evolution problem, we first decompose the $4D$
spacetime in a set of $3D$ spatial hypersurfaces $\Sigma_t$, labeled by a time parameter $t$,
each of them having a $3$-metric $\gamma_{ij}$ given by the space components of
\begin{equation}
\gamma\mndd=g\mndd-n_\mu n_\nu\,.
\end{equation}
Here, $g\mndd$ is the $4D$ spacetime metric, and $n^\mu$ is the unit timelike ($n^\mu n_\mu=-1$) vector normal to the
hypersurfaces $\Sigma_t$.  Thus, we can write the spacetime metric as
\begin{align}
ds^2&=g\mndd dx^\mu dx^\nu \nn\\
&=-(\alpha^2-\beta_i\beta^i)dt^2+2\gamma\ijdd\beta^i dt dx^j+\gamma\ijdd dx^idx^j\,,
\end{align}
where $\alpha$, $\beta^i$ are called lapse function and shift vector, respectively, and contain the information about
how the coordinate system changes from a slice to another. The choice of the lapse and shift, thus, corresponds to the
choice of the foliation of the spacetime. The embedding of the hypersurfaces in the $4D$ spacetime is described by the
extrinsic curvature
\begin{equation}
\label{eq:extrinsic_curv}
K\ijdd=-\frac{1}{2}\mathcal{L}_{\bm n}\gamma\ijdd\,,
\end{equation}
where $\mathcal{L}_{\bm n}=\alpha^{-1}(\partial_t-\mathcal{L}_\beta)$ is the Lie derivative along the unit vector
$n^\mu$. If a scalar field $\Phi$ is also present, we also define its momentum
\begin{equation}
  \label{eq:scalar_momentum}
  \Pi=-\mathcal{L}_{\rm n}\Phi\,.
\end{equation}
Then, all $4D$ quantities such as the Ricci scalar $R$, the Ricci tensor $R\mndd$, etc. can be decomposed in terms
of the $3D$ metric $\gamma_{ij}$. The $3D$ Ricci tensor and Ricci Scalar, $\prescript{3}{}{R}_{ij}$,
$\prescript{3}{}{R}$, the extrinsic curvature and $K\ijdd$, its trace $K=\gamma^{ij}K\ijdd$, and the scalar
field momentum $\Pi$ follow from this $3D$ metric. We denote with $\nabla_\mu$ the covariant derivative with respect to the $4D$ metric $g\mndd$,
and with $D_i$ the covariant derivative with respect to the $3D$ metric $\gamma\ijdd$.

With this decomposition, Einstein's equations (possibly supplemented with the scalar field dynamical equation) lead to
two sets of equations: (i) the evolution equations, which are (with a careful choice of variables) hyperbolic equations
giving the time evolution of the $3D$ quantities; (ii) the constraint equations, which are ellyptic equations which have
to be satisfied at any $3D$ slice $\Sigma_t$. The constraint equations are the  {\it Hamiltonian constraint equation}
\begin{equation}
\label{eq:HGR}
\mathcal{H}^{\rm GR}\equiv\prescript{3}{}{R}+K^2-K\ijdd K\ijuu=0\,,
\end{equation}
and the three {\it momentum constraint equations}
\begin{equation}
\label{eq:MGR}
\mathcal{M}^{\rm GR}_i\equiv D_jK^j_i-D_i K=0\,.
\end{equation}
In presence of matter, the constraint equations are
\begin{align}
  \label{eq:HGRphi}
\mathcal{H}^{\rm GR}&=16\pi\rho\,,\\
  \label{eq:MGRphi}
\mathcal{M}^{\rm GR}_i&=8\pi j_i\,,
\end{align}
where $\rho$ is the energy density of the matter fields, and $j_i$ is its energy-momentum flux. For a scalar
field they are
\begin{align}
\label{eq:rhostt}
\rho&=\frac{1}{2}\Pi^*\Pi+\frac{1}{2}D^i\Phi^*D_i\Phi\,,\\
\label{eq:jstt}
j_i&=\frac{1}{2}\left( \Pi^*D_i\Phi+\Pi D_i\Phi^* \right)\,,
\end{align}
where $\Pi$ is the momentum of the scalar as defined in Eq.\til\eqref{eq:scalar_momentum}. 

In NR one first solves the constraint equations at the initial time $t=t_0$, finding the {\it initial data} on
$\Sigma_{t_0}$ of the system. Then, one solves the evolution equations, finding the spacetime at all times
$t\ge t_0$. Within the CLAP, instead, the evolution of physical initial data is determined through the use of perturbation theory, therefore solving a linearized version of the evolution equations (see below).
\subsection{Initial data for binary black holes in general relativity}\label{subsec:BL}

Solving the constraint equations~\eqref{eq:HGR}-\eqref{eq:MGR} and finding initial data appropriate to study a given
system is a subject of study on its
own\,\cite{Cook:2000vr,alcubierre2008introduction,baumgarte2010numerical,shibata2015numerical}. The original formulation
of the CLAP~\cite{Price:1994pm} used the Misner initial data\,\cite{Misner:1960zz} (a common choice for NR simulations
decades ago) to describe BBH head-on collisions. Subsequently, it was found that the Brill-Lindquist (BL) initial
data\,\cite{Brill:1963yv} are more appropriate for NR simulations, because they have a simpler form and are easier to be
extended to the more realistic non-head-on case (the Bowen-York initial data~\cite{Bowen:1980yu}). Thus, later
applications of the CLAP employ the BL (or Bowen-York) initial
data~\cite{Abrahams:1995wd,Andrade:1996pc,Khanna:1999mh,Sopuerta:2006wj}.  In this article, we shall use BL initial
data and their extensions.

In the case of a head-on collision of two non-rotating compact objects starting from rest, the extrinsic curvature
identically vanishes. Additionaly, considering that in vacuum $\Phi=0$, the constraint
equations\,\eqref{eq:HGR}-\eqref{eq:MGR} reduce to $\prescript{3}{}{R}=0$. The BL three-metric $\gamma_{ij}$ is a
conformally flat solution of this equation describing two BHs. It has the form
\begin{equation} \label{eq:eq:BL_id}
\prescript{3}{}{ds}^2_{\rm BL}=\varphi_{\rm BL}^4 \prescript{3}{}{d\eta}^2=\varphi_{\rm BL}^4 \left( dR^2+R^2d\Omega^2 \right)\,,
\end{equation}
where $d\Omega^2=d\theta^2+d\phi^2\sin^2\theta$ , $\prescript{3}{}{d\eta}^2$ is the flat three-metric, and
$\varphi_{\rm BL}(R,\theta,\phi)$ is the conformal factor. With this choice, the Hamiltonian constraint
$\prescript{3}{}{R}=0$ reduces to $\nabla^2\varphi_{\rm BL}=0$. The BL solution of this equation is
\begin{equation} \label{eq:BL_id_newtonian}
\varphi_{\rm BL}=1+\frac{m_1}{2|{\bm R}-{\bm R_1}|}+\frac{m_2}{2|{\bm R}-{\bm R_2}|}\,,
\end{equation}
where the vector ${\bm R}$ is the position vector in the flat three-space, and ${\bm R_i}$ is the position of the $i$-th
BH; $M=m_1+m_2$ is the ADM mass of the entire spacetime. Note that $m_1$ and $m_2$ are not the ADM masses of the two
BHs, which are~\cite{Brill:1963yv}
\begin{equation}\label{eq:bare_masses}
M_1=m_1\left(1+\frac{m_2}{2d}\right)\,,~~M_2=m_2\left(1+\frac{m_1}{2d}\right)\,,
\end{equation}
and $d=|{\bm R_2}-{\bm R_1}|$. We choose the origin of the coordinate system in the center of mass: $m_1{\bm
  R_1}+m_2{\bm R_2}={\bm 0}$. In the region of spacetime where $R>R_1,R_2$, the conformal factor can be expanded in
terms of Legendre polynomials:
\begin{equation}
\varphi_{\rm BL}=1+\frac{M}{2R}+\sum_{\ell=1}^{\infty}\xi_\ell\left( \frac{M}{R} \right)^{\ell+1} P_\ell\left(\cos\theta\right)\,.
\end{equation}
The coefficients $\xi_\ell$ are given by~\cite{Andrade:1996pc}
\begin{equation}
\xi_\ell=\left(\frac{R_1}{M}\right)^{\ell}\frac{m_1}{2M}+\left(\frac{R_2}{M}\right)^{\ell}\frac{m_2}{2M}\,.\label{eq:defxi0}
\end{equation}
In the following we shall consider equal-mass binary systems; thus $m_1=m_2=M/2$. By choosing the $Z$-axis aligned with
the motion, ${\bm R_{1/2}}=(0,0,\pm Z_0)$ with $Z_0=d/2$. Thus $\xi_1=0$, and Eq.~\eqref{eq:defxi0} reduces to
\begin{equation} \label{eq:xi_M1eqM2}
\xi_\ell=\frac{1}{2}\left(\frac{Z_0}{M}\right)^{\ell}\,,\;\;\; \text{for }\ell=2,4,6,\dots \, .
\end{equation}
When the coefficients $\xi_\ell$ vanish, $\varphi_{\rm BL}=1+M/(2R)$ and the metric~\eqref{eq:eq:BL_id} describes the
$t=const.$ slices of the Schwarzschild's background. Indeed, by defining the Schwarzschild radial coordinate $r$ in terms
of the isotropic radial coordinate $R$ by
\begin{equation}
\label{eq:R_isotropic_to_Schw}
R=\frac{1}{4}\left(\sqrt{r}+\sqrt{r-2M}\right)^2\,,
\end{equation}
Eq.~\eqref{eq:eq:BL_id} gives
\begin{equation}
\prescript{3}{}{ds}^2_{\rm BL}=\mathcal{F}_{\rm BL}^4\left(f^{-1}dr^2+r^2d\Omega^2\right)\,,
\end{equation}
where
\begin{align}
\label{eq:f_of_Schwa}
f(r)&=1-\frac{2M}{r}\,,\\
\mathcal{F}_{\rm BL}&\equiv\varphi_{\rm BL}\left( R,\theta\right) \left( 1+M/2R\right)^{-1}\,.
\end{align}
If $\varphi_{\rm BL}=1+M/(2R)$,
$\mathcal{F}_{\rm BL}=1$ and by defining the $4D$ spacetime metric as 
\begin{equation}
\label{eq:pert_Schw_CLAP_GR_BL}
ds^2=-f dt^2+\prescript{3}{}{ds}^2_{\rm BL}
\end{equation}
(corresponding to an appropriate choice of the shift vector and lapse function, i.e. to an appropriate gauge choice), it
coincides with the Schwarzschild geometry. If, instead, the coefficients $\xi_\ell$ are small but non-vanishing, the
metric~\eqref{eq:eq:BL_id} is a perturbation of Schwarzschild's three-metric, and Eq.~\eqref{eq:pert_Schw_CLAP_GR_BL} is
a perturbation of Schwarzschild's spacetime.
Therefore the BL three-metric, in the coordinates $(r,\theta,\phi)$, has the form
\begin{align} \label{eq:dsBL_M1M2z0}
\prescript{3}{}{ds}^2_{\rm BL}&= \left( 1+\frac{1}{1+\frac{M}{2R}} \sum_{\ell=2,4,\dots}^{\infty}
\xi_\ell \left(\frac{M}{R}\right)^{\ell+1}P_\ell\left(\cos\theta\right)\right)^4\nn\\
&\times \left(f^{-1} dr^2+r^2d\Omega^2\right)\,.
\end{align}
If $\xi_\ell\ll1$ we can linearize in the parameters $\xi_\ell$, and Eq.~\eqref{eq:dsBL_M1M2z0} gives:
\begin{align}
\label{eq:dsBL_pert}
\prescript{3}{}{ds}^2_{\rm BL}&= \left( 1+\frac{4}{1+\frac{M}{2R}}\sum_{\ell=2,4,\dots}^{\infty}
\xi_\ell\left(\frac{M}{R}\right)^{\ell+1} P_\ell\left(\cos\theta\right)\right)\nn\\
&\times\left(f^{-1} dr^2+r^2d\Omega^2\right)\,.
\end{align}
The parameter $Z_0$ in Eq.~\eqref{eq:xi_M1eqM2} describes the (initial) distance
between the BHs in the isotropic frame. This quantity does not have a direct physical interpretation; thus, several
authors characterize the initial BH separation with the proper distance $L$ between the apparent horizons of the two
BHs, given by~\cite{Andrade:1996pc}
\begin{equation}
  L=\int_{Z_1}^{Z_2}\left[1+\frac{M}{4}\left(\frac{1}{Z_0+Z}+\frac{1}{Z_0-Z}\right)\right]^2dZ\,.\label{eq:defL}
\end{equation}
The extrema $Z_1$, $Z_2$ are the intersections of the apparent horizons of the two BHs with the $Z$-axis, and can be
found by numerical integration of the equations describing the apparent
horizons~\cite{bishop1982closed,bishop1984horizons}; the procedure to compute these quantities is described in detail,
for instance, in~\cite{Sopuerta:2006wj}. This computation shows that, for example, $L=3M$ for $Z_0\simeq0.5M$, $L=3.5M$
for $Z_0\simeq0.7M$, $L=4M$ for $Z_0\simeq0.85M$.

A comparison with NR computations in the head-on case shows that the CLAP is accurate, in the equal-mass case, for
$L\lesssim4M$ (see~\cite{Gleiser:1998rw} and the discussion in~\cite{Sopuerta:2006wj}), corresponding to
$Z_0\lesssim0.85M$ and thus $\xi_2\lesssim0.36$; as noted in~\cite{Price:1994pm}, it is remarkable that the agreement
extends far beyond the region $\xi_\ell\ll1$ in which the CLAP is expected to be applicable. Therefore, we shall apply
Eq.~\eqref{eq:dsBL_pert} and, more generally, perturbation theory, also to initial separations for which the condition
$\xi_\ell\ll1$ is only marginally satisfied.
%
\subsection{Perturbations and their time evolution}\label{subsec:recast_BH}
The metric in Eq.~\eqref{eq:pert_Schw_CLAP_GR_BL} describes the Schwarzshild spacetime, which we consider as the
background, with a perturbation with even parity, at a fixed ``initial'' time $t=t_0$. Thus, it can be recast in the
form
\begin{equation}
g_{\mu\nu}=g^{(0)}_{\mu\nu}+h_{\mu\nu}\,,
\end{equation}
where
\begin{equation}\label{eq:Schwarzschild_metric}
g^{(0)}_{\mu\nu}={\rm  diag}(-f,f^{-1},r^2,r^2\sin^2\theta)
\end{equation}
is the Schwarzschild spacetime (with $f$ defined in Eq.\til\eqref{eq:f_of_Schwa}), and $h_{\mu\nu}(t,r,\theta,\phi)$ is
the perturbation, whose evolution in the wave zone $r\gg M$ can be described in terms of a scalar function called the
Zerilli function~\cite{Zerilli:1971wd}. The original definition of the Zerilli function is appropriate for the study of
oscillating solutions; in this case, we want to study the evolution of a given set of initial data at the time $t=t_0$;
thus, it is more appropriate the definition of the Zerilli function in terms of gauge-invariant quantities given
in~\cite{Moncrief:1974am,Cunningham:1979px}.

We first note that the only non-vanishing perturbations in the
metric~\eqref{eq:pert_Schw_CLAP_GR_BL}-\eqref{eq:dsBL_pert} are 
\begin{align}
  h_{rr}(r,\theta)&=f^{-1}\sum_{\ell=2,4,\dots}g_\ell(r) \xi_\ell P_\ell(\cos\theta)\,,\nonumber\\
  h_{\theta\theta}(r,\theta)&=\frac{h_{\phi\phi}}{\sin^2\theta}=r^2\sum_{\ell=2,4,\dots}g_\ell(r)\xi_\ell P_\ell(\cos\theta)\,,
\label{eq:BLrecast0}
\end{align}
where
\begin{equation}
g_\ell=4\left(1+\frac{M}{2R}\right)^{-1}\frac{M^{\ell+1}}{R^{\ell+1}}\,.
\end{equation}
Using the notation of~\cite{Cunningham:1979px} for polar parity, axially symmetric perturbations,
Eqs.~\eqref{eq:BLrecast0} correspond to the perturbation functions $H_2^\ell(t,r)$, $K^\ell(t,r)$ given by 
\begin{align}
h_{rr}(t,r,\theta)&=f^{-1}\sum_\ell H_2^\ell(t,r)P_\ell(\cos\theta)\,,\nonumber\\
h_{\theta\theta}(t,r,\theta)&=r^2\sum_\ell K^\ell(t,r)P_\ell(\cos\theta)\,,\label{eq:BLrecast}
\end{align}
which implies that at the initial time $t=t_0$,
\begin{equation}\label{eq:H2_Kl_equality}
H_2^\ell(t_0,r)=K^\ell(t_0,r)=g_\ell\xi_\ell,~~{\text{with }}\ell=2,4,6,\dots\,,
\end{equation}
while the other perturbation functions ($H_0^\ell,H_1^\ell$) and ($G^\ell,h_0^\ell,h_1^\ell$) identically vanish at $t=t_0$.
Since the leading contribution comes from the quadrupole perturbations, we shall consider the $\ell=2$ contribution only. 
Then, following~\cite{Cunningham:1979px}, we define the $\ell=2$ Zerilli function at $t=t_0$,
\begin{equation}
  \psi(t_0,r)=\sqrt{\frac{4\pi}{5}}{\lambda}^{-1}Q(r)\xi_2\,,\label{eq:zerillit0}
\end{equation}
where $\lambda=1+3M/(2r)$ and (leaving implicit the index $\ell=2$ and the arguments of $H_2(t_0,r)$ and $K(t_0,r)$)
\begin{align}
  Q(r)&=\frac{2rf^2}{\xi_2}
  \left[\frac{H_2}{f}-\frac{1}{\sqrt{f}}\frac{d}{dr}\frac{rK}{\sqrt{f}}\right]+\frac{6r}{\xi_2}K\nonumber\\
 &= 2r f^2\left[\frac{g}{f}    -\frac{1}{\sqrt{f}}\frac{d}{dr}\frac{r g}{\sqrt{f}}\right]+6rg\label{eq:defQ}\,,
\end{align}
with $g=4\left(1+M/(2R)\right)^{-1}M^3/R^3$.
The perturbation equations can be written as a wave equation for the Zerilli function,
\begin{equation}
-\frac{\partial^2\psi}{\partial t^2}+\frac{\partial^2\psi}{\partial r_*^2}-V_Z\psi=0\label{eq:zer}\,,
\end{equation}
where $r_*=r+2M\log\left|\frac{r}{2M}-1\right|$ is the tortoise coordinate, and
\begin{equation}\label{eq:zer_potential}
  V_Z=f\left\{\frac{1}{{\lambda}^2}\left[\frac{9M^3}{2r^5}-\frac{3M}{r^3}\left(1-\frac{3M}{r}\right)\right]
  +\frac{6}{r^2\lambda}\right\}
\end{equation}
is the Zerilli potential. Eq.~\eqref{eq:zer} is called the Zerilli equation~\cite{Zerilli:1971wd}.

In most implementations of the CLAP, the Zerilli equation~\eqref{eq:zer} with initial condition~\eqref{eq:zerillit0} is solved in the time domain. Alternatively, it can be solved in the frequency domain. Indeed, it can be shown (see
e.g.~\cite{Lousto:1996sx}) that if the function $\psi(t,r_*)$ is the solution of the Zerilli equation~\eqref{eq:zer} with
initial conditions~\eqref{eq:zerillit0}  and $\dot\psi(t_0,r_*)=0$ (i.e., assuming stationarity of the initial data), then,
choosing the time coordinate such that $t_0=0$, 
the Laplace transform
\begin{equation}\label{eq:laplace_psi}
  \tilde\psi(\omega,r_*)=\int_{0}^\infty dt\psi(t,r_*)e^{i\omega t}\,,
\end{equation}
is the solution of the Zerilli equation with source
\begin{equation}
\frac{\partial^2\tilde\psi}{\partial r_*^2}+(\omega^2-V_Z)\tilde\psi=S\,,\label{eq:zerS}
\end{equation}
where
\begin{equation}\label{eq:zer_source}
S(\omega,r_*)=i\omega\psi(t=0,r_*)\,,
\end{equation}
and boundary conditions of ingoing wave at the horizon, outgoing wave at infinity (see Eq.~\eqref{eq:bcsomm1} in
Appendix~\ref{app:solveZerBH}).
Finally, the total energy emitted in GWs in the collision is given by~\cite{Price:1994pm}
\begin{equation}
E=\frac{1}{384\pi}\int_0^\infty \left\lvert \frac{\partial \psi}{\partial t}\right\rvert^2 dt\,.\label{total_E}
\end{equation}
We have solved Eq.~\eqref{eq:zerS} using two different approaches (see Appendix~\ref{app:solveZerBH} for details),
reproducing in both cases the results in the literature for head-on BBH collisions with the CLAP
approach~\cite{Price:1994pm}. The total energy obtained with our computation agrees with that of
Ref.~\cite{Price:1994pm} within one percent.

In this article we only consider head-on collisions of non-rotating, equal mass compact objects starting from rest as
in~\cite{Price:1994pm}; however, it is worth mentioning that the CLAP approach has been extended to more realistic
setups, considering BHs with initial velocity~\cite{Baker:1996bt}, unequal masses~\cite{Andrade:1996pc} and non-head-on
binary inspirals~\cite{Khanna:1999mh}.

\section{Extreme Compact Objects}\label{sec:BH_mimickers}
We shall now apply the CLAP to describe the head-on collision of two non-rotating, equal mass ECOs. We model an ECO (see
e.g.~\cite{Cardoso:2019rvt} and references therein) as a spherically symmetric compact body with mass $M$ and a surface at $r=r_0$,
\begin{equation}\label{eq:ECO_radius}
  r_0=2M(1+\epsilon)\,,
\end{equation}
with $\epsilon\ll1$.
The above parametrization for the ECO's surface well describes ECO models for which $\epsilon\ll 1$, i.e. gravastars, wormholes, fuzzballs etc.~\cite{Cardoso:2019rvt}. In the following we restrict to ultra-compact bodies whose surface can be described by Eq.\til\eqref{eq:ECO_radius}. This means that the dynamical processes involving such ECOs resemble those of BHs (with due differences, as we will see shortly) rather than stellar objects such as neutron stars for which $\epsilon\sim\mathcal{O}\left(1\right)$).

A fundamental difference between the surface of an ECO and the horizon of a BH is that an incoming wave is partially
reflected by the ECO surface, while it is totally absorbed by a BH horizon. Thus, given a (scalar, gravitational, etc.)
test field $\phi(t,r_*)\sim\tilde\phi(r_*)e^{i\omega t}$, its boundary condition near the surface $r=r_0$ is
\begin{equation}
  \tilde\phi(r_*)\sim e^{-i\omega r_*}+\mathfrak{R}e^{i\omega r_*}\,.\label{bceco0}
\end{equation}
The parameter $\mathfrak{R}$, which in general depends on the frequency $\omega$ and on the spin of the field, is called
{\it reflectivity coefficient} of the ECO.
\subsection{Initial data for extreme compact objects}\label{subsec:BL-ECO}
We shall first introduce the spacetime metric of a single, isolated, spherically symmetric ECO; then, using the
procedure of Brill and Lindquist discussed in Sec.~\ref{subsec:BL}, we shall define initial data corresponding to two
ECOs starting from rest (this approach can naturally be extended to the case of ECOs with initial momentum, using the
procedure of Bowen and York~\cite{Bowen:1980yu}).

The GW signal produced by the collision of ECOs will consist of waves caused by excitations of the exterior
spacetime~\cite{Vishveshwara:1970zz,Berti:2009kk}, as well as contributions which probe the interior of the
object~\cite{Cardoso:2016rao,Cardoso:2016oxy,Cardoso:2017cqb,Abedi:2016hgu,Mark:2017dnq,Testa:2018bzd,Maggio:2018ivz}. For
a wide class of asymptotically flat ECOs, the lapse in its interior is very small, leading to large delays in signals which have probed the
interior geometry~\cite{Ferrari:2000sr,Cardoso:2017cqb,Cardoso:2019rvt}. Thus, in the following we ``freeze'' the inner
region, which means that in practice we cut it off from our domain, and all the information about the interior is
replaced by boundary conditions at the surface. 
  
Let us first consider a single, isolated (spherically symmetric) ECO. Due to Birkhoff's theorem, the exterior of the ECO
$r>r_0$ is described by Schwarzschild's geometry (Eq.\til\eqref{eq:Schwarzschild_metric})\,\footnote{Since
  $\epsilon\ll1$, $r_0<3M$ and thus the light ring $r=3M$, where null circular geodesics are defined, lies in the
  exterior of the ECO. This plays a crucial role in the GW emission, as we discuss below.}. We define the isotropic
radial coordinate as in Schwarzschild metric~\eqref{eq:R_isotropic_to_Schw}:
\begin{equation}
R=\frac{1}{4}\left( \sqrt{r}+\sqrt{r-2M}\right)^2\,.\label{eq:R_isotropic_to_ECO}
\end{equation}
In terms of the coordinate $R$, the location of the surface is $R_0=R(r_0)=M/2(1-2\sqrt{\epsilon})$.
The interior of the ECO depends on the profile of its energy density $\rho(R)$. As discussed above, we do not need to
model the interior because, under our assumptions, the emitted GW signal does not depend on the structure of the
interior; it only depends on the location of the surface and on its properties. We stress that ``freezing'' the internal degrees of freedom of the ECOs is a standard approximation in the literature of ECO modeling~\cite{Cardoso:2019rvt}.

Let us now consider two spherically symmetric ECOs, with equal masses and total ADM mass $M$,
initially at rest.  We perform a $3+1$ decomposition of the spacetime, as in Sec.~\ref{sec:3+1}. The three-metric of
each slice is solution of the Hamiltonian constraint equation~\eqref{eq:HGRphi}:
\begin{equation} 
\mathcal{H}^{\rm GR}= \prescript{3}{}{R}+K^2-K\ijdd K\ijuu=16 \pi \rho\,.\label{eq:H_ECO}
\end{equation}
Since the ECOs are initially at rest, the initial extrinsic curvature vanishes and Eq.~\eqref{eq:H_ECO} reduces to
$\prescript{3}{}{R}= 16 \pi \rho$. In the exterior of both ECOs, $\rho=0$ and the metric can be written in the BL 
form, Eq.~\eqref{eq:dsBL_pert}:
\begin{align}
\label{eq:dsBL_pert_ECO}
\prescript{3}{}{ds}^2_{\rm BL-ECO}&= \left( 1+\frac{4}{1+\frac{M}{2R}} \sum_{\ell=2,4,\dots}^{\infty}
\xi_\ell P_\ell\left(\cos\theta\right)\right)\nn\\
&\times\left(f^{-1} dr^2+r^2d\Omega^2\right)\,,
\end{align}
with $\xi_\ell=\left(Z_0/M\right)^\ell/2$ as in Eq.~\eqref{eq:xi_M1eqM2}. In the isotropic frame, the two ECOs move
along the $z$-axis and are located at $R=\pm Z_0$, with $0<Z_0\lesssim 0.85M$ (see also Sec.~\ref{subsec:BL}). We call
Eq.~\eqref{eq:dsBL_pert_ECO} the BL-ECO initial data.

We remark that although the BL initial data for BBHs in Eq.~\eqref{eq:dsBL_pert} and the BL-ECO initial data for binary
ECOs in Eq.~\eqref{eq:dsBL_pert_ECO} are formally identical, the former are defined outside the BH horizon, while the
latter are defined outside the ECO surface. This difference, as we shall show below, leads to a difference in the GW
emission. Note also that, as discussed above, the region inside the surfaces of the ECOs does not contribute to the GW signal, by assumption.
\subsection{Collision of extreme compact objects}\label{subsec:ECO-collision}
%
We shall now recast the BL-ECO initial data in Eq.~\eqref{eq:dsBL_pert_ECO} as a perturbation of a single object,
\begin{equation}
\label{eq:eco_exp}
g_{\mu\nu}=g^{(0)}_{\mu\nu}+h_{\mu\nu}\,.
\end{equation}
Then, we shall evolve them using the tools of perturbation theory.

Since we are not dealing with vacuum spacetimes, the outcome of the ECO collision depends, in principle, on the
features of the ECOs, i.e. on their internal structure. Two outcomes are possible: either the collision leads to a
single BH (which, in the head-on collision of non-rotating ECOs, is described by the Schwarzschild metric), or it leads to an
ECO. In the former case the initial data should be recast as a perturbation of a Schwarzschild solution; in the latter,
as a perturbation of an ECO. We shall treat these two cases separately.

\subsubsection{Formation of a black hole}
If the final object is a Schwarzschild BH, the background metric $g^{(0)}_{\mu\nu}$ in Eq.~\eqref{eq:eco_exp} is
Schwarzschild's metric. The procedure of recasting the BL-ECO initial data~\eqref{eq:dsBL_pert_ECO} as a perturbation of
Schwarzschild metric is formally equivalent to the derivation in Sec.~\ref{subsec:recast_BH}, leading to an $\ell=2$
Zerilli function at the initial time $t=t_0$ given by Eqs.~\eqref{eq:zerillit0}-\eqref{eq:defQ}.

Since the background is Schwarzschild's metric, the Zerilli equation~\eqref{eq:zer} has to be solved with the same
boundary condition as in the BBH case. Therefore, the gravitational waveform emitted in the
collision of two ECOs into a Schwarzschild BH, computed with the CLAP (and neglecting the effect of perturbations inside
the ECO's surface) coincides with the waveform emitted in a BBH collision. \footnote{Note that this does not imply that generic theories of gravity yield the same dynamics as GR, if the final merger product is a BH. The implication is that, if the initial state is parametrically close to a BH, and the final state is a Kerr BH, then the approach to the final state is identical within GR.}

\subsubsection{Formation of an extreme compact object}
If the final object is an ECO, the background metric $g^{(0)}_{\mu\nu}$ is Schwarzchild metric only for $r>r_0$. 

It seems extraordinary that two compact objects coalesce and are prevented to collapse completely to a BH. However, we intend on mimicking the effects of a theory where BHs are absent (via, for example, high-energy phenomena). We may assume, for example, that a theory of quantum gravity forbids the existence of horizons via Planck-scale physics. We do not wish to build such a theory but merely to investigate some of its consequences. We do assume that departures from GR occur only close to the Schwarzschild radii.

We remark that, as discussed in the Introduction, this
computation is based on strong assumptions (the bodies collide head-on, they have no spin, and most importantly the
``initial data'' are stationary, although they describe the late stage just before the merger). Therefore, our results should
only be considered as a order-of-magnitude estimates. The computation itself should be considered as the first step in
the modeling of coalescences with ECOs using the CLAP. These approximations will be relaxed in future publications.


The procedure of recasting the BL-ECO initial data~\eqref{eq:dsBL_pert_ECO} as a perturbation of the ECO
metric follows the derivation in Sec.~\ref{subsec:recast_BH}, leading to the Zerilli function at the initial time
$t=t_0$ given in Eqs.~\eqref{eq:zerillit0} and \eqref{eq:defQ}. However, the Zerilli function $\psi(t,r_*)$ is only defined
for $r_*>r_{0*}=r_0+2M\log|r_0/(2M)-1|$, and it satisfies boundary conditions different from those of a BH.

Using the Laplace transform approach discussed in Sec.~\ref{subsec:recast_BH}, one finds the Zerilli equation with source
given in Eqs.~\eqref{eq:zerS} and \eqref{eq:zer_source}. Although the equation is the same as for BBH collisions, the
boundary conditions are different: since the surface is partially reflecting, they are (see Eq.~\eqref{bceco0})
\begin{align} \label{eq:BC_ECO}
  \tilde{\psi}&\sim e^{-i \omega  \left( r_*- r_{0*}\right)}+ \mathfrak{R}\,
  e^{i \omega \left( r_*- r_{0*}\right)}~~~~~~(r_*\to r_{0*})\nn\\
\tilde{\psi}&\sim  e^{i \omega r_*} \hspace{4.2cm} (r_*\to\infty)\,,
\end{align}
where $\mathfrak{R}$ is the reflectivity coefficient of the ECO\,\footnote{ In general
  $\mathfrak{R}$ is a function of $\omega$; for simplicity, we shall assume it to be a constant.}.

We solved the Zerilli equation~\eqref{eq:zerS} with boundary conditions\,\eqref{eq:BC_ECO} using a shooting method 
(see Appendix~\ref{app:solveZerECO}), for different values of $\epsilon$ and of
$\mathfrak{R}$, finding the gravitational waveform emitted in the merger and ringdown phases of the collision. Since we only consider the leading quadrupolar contribution, the Zerilli function is proportional to $\xi_2$, Eq.~\eqref{eq:zerillit0}, and thus to $Z_0^2$. Moreover, it has (in geometric units) the dimensions of length and thus,
since the only dimensionful quantity characterizing the ECO is its ADM mass $M$, $\psi(t,r_*)\propto Z_0^2/M$.
\subsubsection{Results of the numerical integration}

\begin{figure}[ht]
\includegraphics[width=8cm,keepaspectratio]{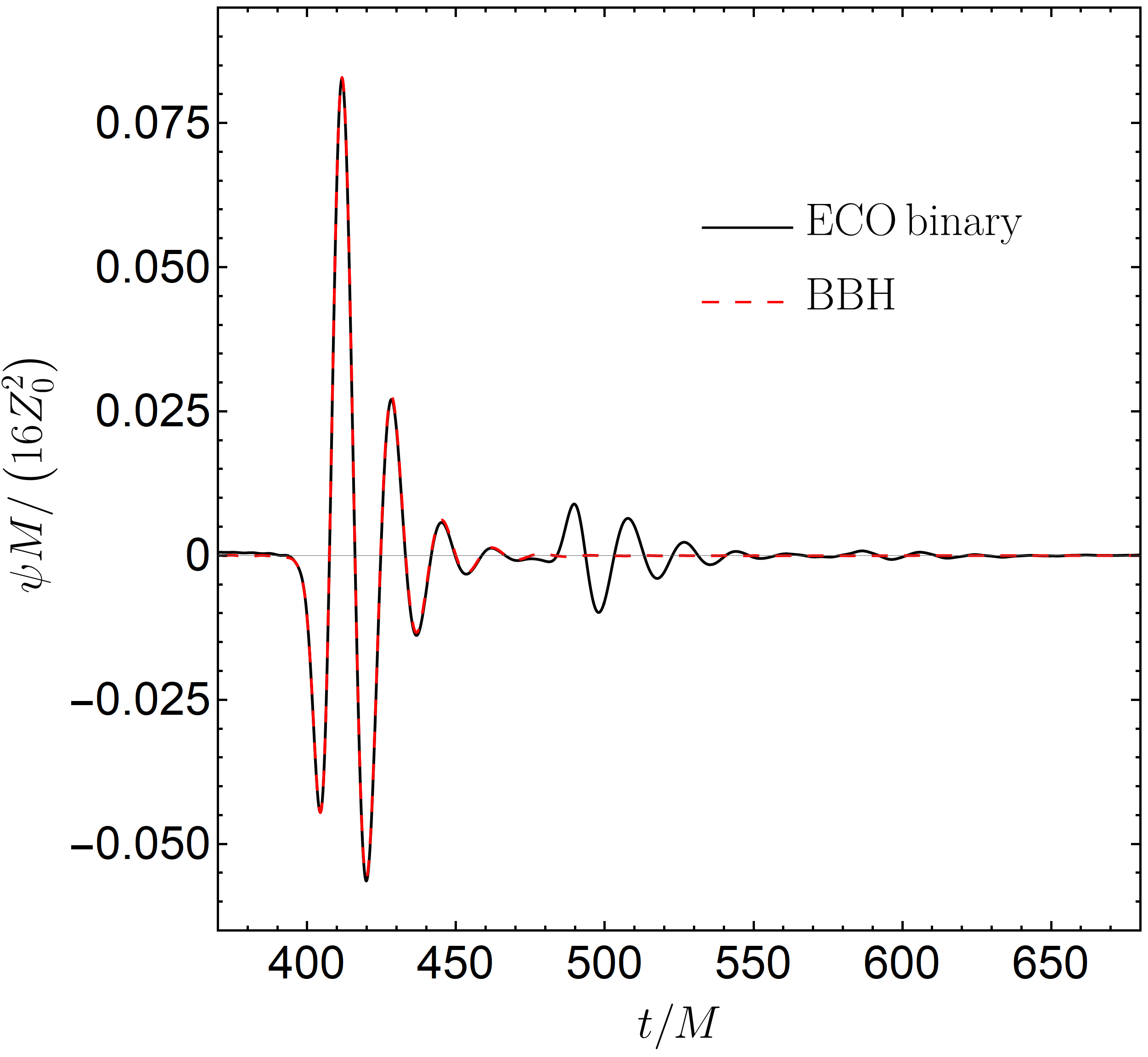}
\caption{Gravitational waveform for the head-on collision of two equal-mass, spherically symmetric BHs/ECOs starting
  from rest, evaluated at the extraction radius $r_*^{extr}=400\,M$, for two possible outcomes. If
  the final object is a BH, the waveform is identical to that resulting from the collision of two BHs: a sharp burst
  followed by ringdown (caused by the relaxation of the light ring). When the final product is an ECO, we observe
  a similar initial stage, followed at late times by echoes of the initial
  burst~\cite{Cardoso:2016oxy,Cardoso:2019rvt}. Here, the final object is taken to have a reflectivity of
  $\mathfrak{R}=0.1$ and a surface $r_0=2M(1+\epsilon)$ with $\epsilon=10^{-10}$.}
\label{fig:BBHvsBECO}
\end{figure}
\begin{center}
\begin{figure}
\includegraphics[width=8.5cm,keepaspectratio]{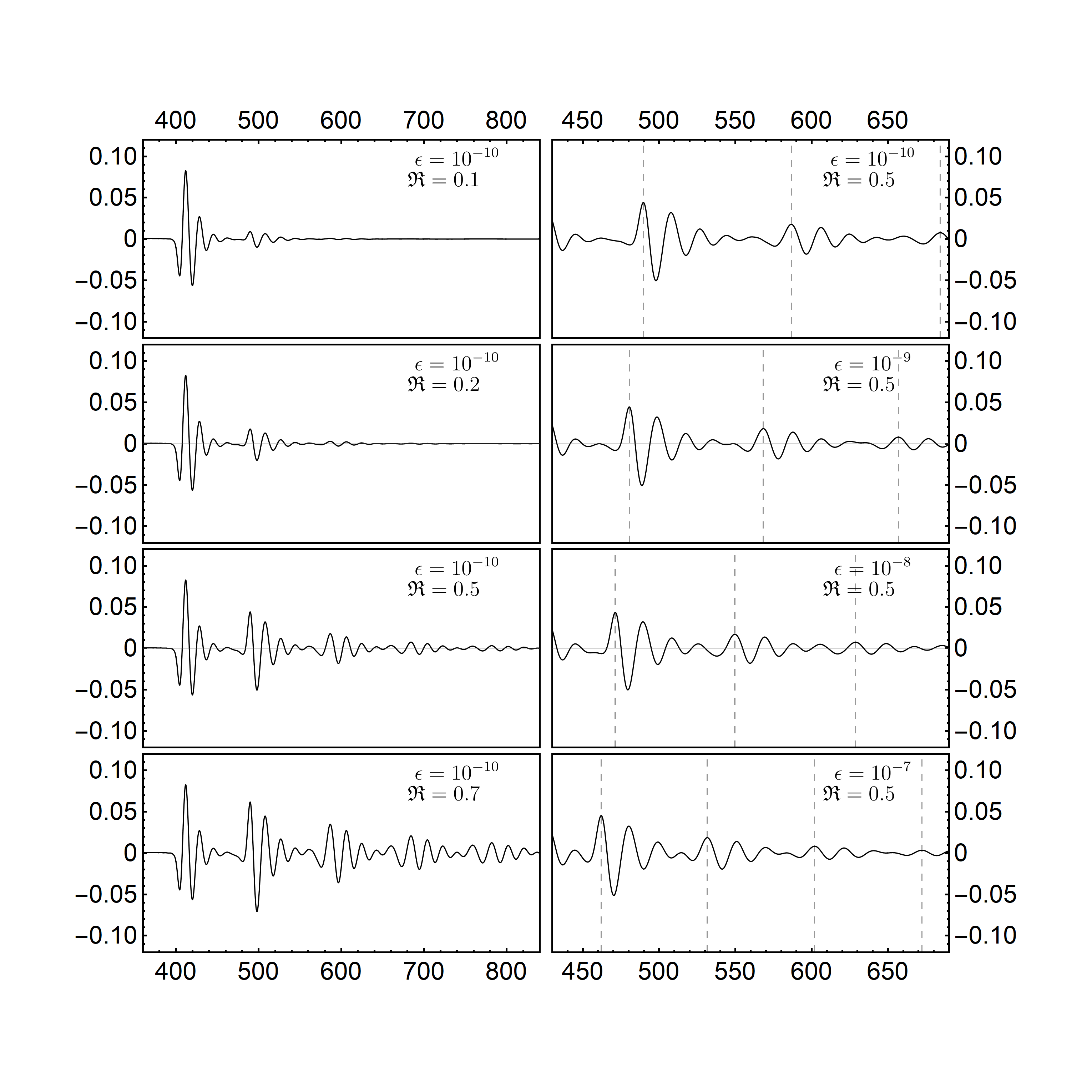}
\caption{Gravitational waveforms for head-on collisions of two equal-mass, spherically symmetric ECOs starting from rest,
  leading to an ECO, for different values of the surface parameter $\epsilon$ and of the reflectivity $\mathfrak{R}$,
  evaluated at the extraction radius $r_*^{extr}=400\,M$ and rescaled by $16Z_0^2/M$. The vertical dashed lines correspond to the
  peak value of each echo.
  The amplitude of the echoes increases with $\mathfrak{R}$, while the time delay between echoes (i.e. between the dashed
  lines) decreases when $\epsilon$ increases.}
\label{fig:Echoes_plot}
\end{figure}
\end{center}

Figure~\ref{fig:BBHvsBECO} shows the Zerilli quadrupolar waveform generated by the head-on collision of two spherically
symmetric compact objects starting from rest, evaluated at the extraction radius $r_*^{\rm extr}=400\,M$. The waveforms
are rescaled by $16Z_0^2/M$, and thus they do not depend neither on $Z_0$ nor on $M$. We compare processes leading to a
Schwarzschild BH (dashed curve) and to an ECO with $\epsilon=10^{-10}$ and $\mathfrak{R}=0.1$ (solid line).  The signals
are identical at early times while they differ at late times, since the ECO collision signal is characterized by a
series of ``echoes'', which are a characteristic feature of GW signals from ECOs (see
e.g.~\cite{Cardoso:2016rao,Cardoso:2016oxy,Mark:2017dnq,Cardoso:2017cqb,Correia:2018apm,Cardoso:2019rvt}). We stress
again that, in this approximation, the signal depends on the nature of the final object, but it does not depend on
whether the colliding objects are BHs or ECOs.

A simple interpretation of the echo structure is the following. The light ring excitation by the initial data is followed by its
``relaxation''. Since both BHs and ECOs have equivalent geometries close to the light ring, they both relax in the same
way~\cite{Cardoso:2016rao,Cardoso:2016oxy,Cardoso:2017cqb,Cardoso:2019rvt}.  This relaxation produces an outgoing wave,
which corresponds to the early part and ringdown of the signal.  But there's also an ``ingoing'' wave which interacts
with the ECO via boundary conditions. Due to a nonzero reflectivity, the ingoing pulse is partly reflected back, and
interacts with the light ring again. The process continues and produces a sequence of distorted copies of the original
burst, i.e. the echoes. 

Figure~\ref{fig:Echoes_plot} displays the dependence of the Zerilli waveform on the ECO parameters $\epsilon$ and $\mathfrak{R}$. We can
see that the echo delay increases with $\epsilon$, while the amplitude of the echoes is larger for larger values of the
reflectivity $\mathfrak{R}$.

We find that the time separation between echoes is well described (with a relative error smaller than $2\%$) by the
following analytical fit,
\begin{equation}
\Delta t_{\rm ECO}\sim 4.3 \left\lvert \log\epsilon \right\rvert M\,,\label{eq:echo_delay}
\end{equation}
for $10^{-10}<\epsilon <10^{-6}$. This logarithmic behaviour is consistent with our knowledge of echoes in the GW signals~\cite{Cardoso:2019rvt}. We only solved the Zerilli equation for
$\epsilon\ge10^{-10}$; for smaller values of $\epsilon$, our numerical approach loses accuracy; however, since this
limitation is of computational nature only, we expect the fit~\eqref{eq:echo_delay} to hold for smaller values of
$\epsilon$ as well.

\begin{figure}
\includegraphics[width=8cm,keepaspectratio]{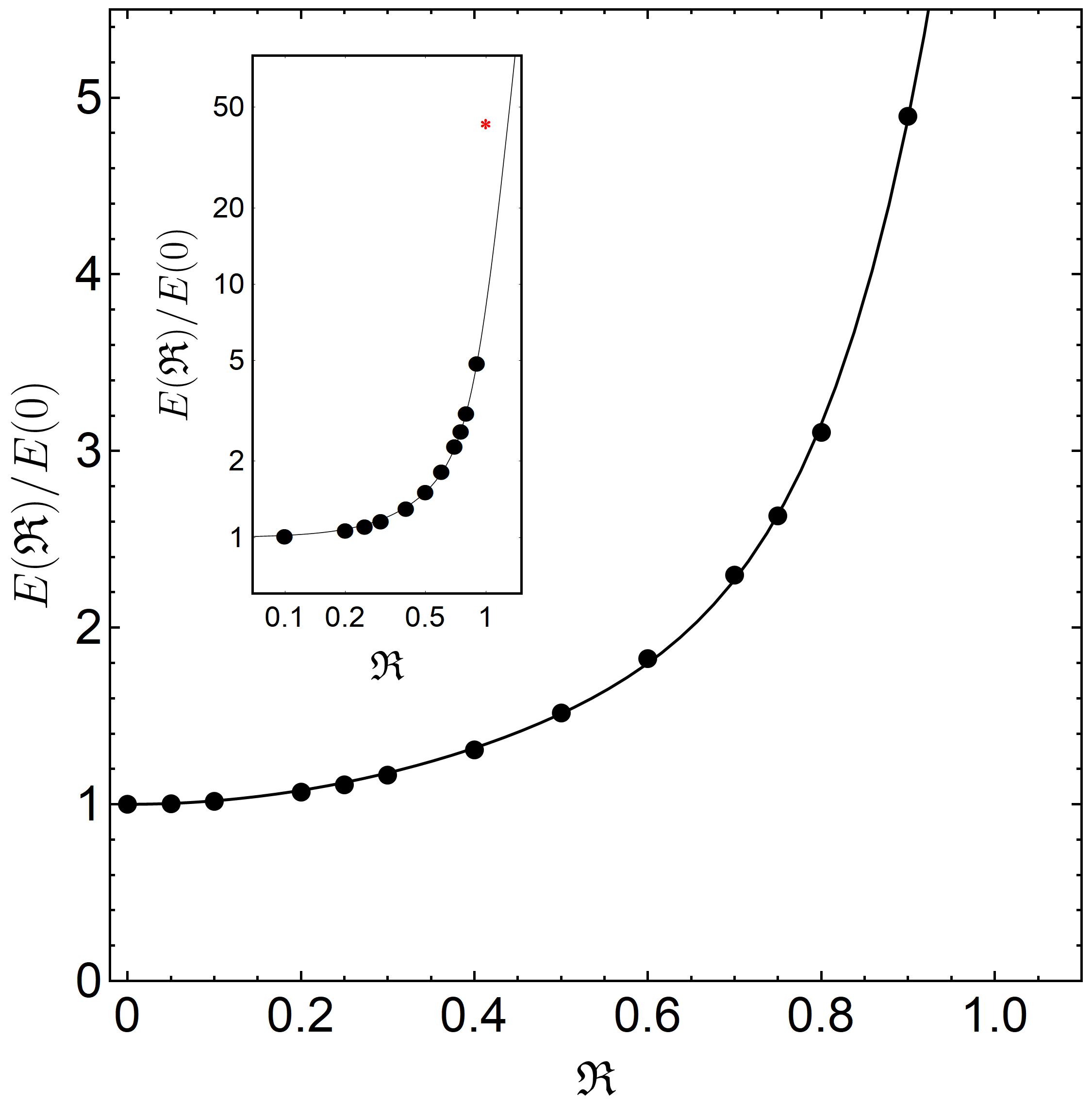}
\caption{Total energy emitted in the head-on collision from rest of two equal-mass, spherically symmetric compact objects. Fixing the ECOs/BHs initial separation $Z_0$, the ratio of energy emitted is independent of such quantity. The final object is assumed to be an ECO with $\epsilon=10^{-10}$ and the results are shown as a function of the reflectivity $\mathfrak{R}$. {\bf Inset:} log-log scale of the same results, including $\mathfrak{R}=1$. This also corresponds to the formation of a wormhole~\cite{Cardoso:2016rao} (see main text).}
\label{fig:E_vs_reflectivity}
\end{figure}

Finally, we compared the total energy $E$ emitted in GWs during a compact object collision leading to an ECO, with that emitted in
a BBH collision. It can be computed in terms of the Zerilli function from Eq.~\eqref{total_E}. We find that, for
$\epsilon<10^{-3}$, the energy radiated is very weakly dependent on $\epsilon$, while it increases with the reflectivity
$\mathfrak{R}$. This is due to the fact that the presence of echoes contributes to the energy loss of the system in GWs.

In Fig.~\ref{fig:E_vs_reflectivity} we show the total energy $E$ emitted in GWs as a function of $\mathfrak{R}$,
normalized by the energy emitted for $\mathfrak{R}=0$ (which coincides with the result for a BBH collision~\cite{Price:1994pm}). 
This shows which part of the emitted energy is due to echoes. The function $E(\mathfrak{R})$ can be
described by the following fit (accurate within $1.5\%$ in the range $0\le \mathfrak{R}\le 0.8$):
\begin{equation}
\frac{E}{M}\approx 10^{-6}\left(6.14+  \mathfrak{R}^{2} (1.29+3.26 \mathfrak{R}^{6})\right)\frac{256Z_0^4}{M^4}\,.\label{eq:rad_fit}
\end{equation}

One can use the CLAP formalism to understand, in particular, the radiation given away during the formation of
wormholes~\cite{Morris:1988tu,Visser:1995cc}. A class of these objects can be considered as ECOs with
$\mathfrak{R}=1$, when the ``throat'' is made of a rigid shell~\cite{Cardoso:2016rao}. This corresponds to taking
$\mathfrak{R}=1$ in the above framework.  The inset of Fig.~\ref{fig:E_vs_reflectivity} shows the total energy computed
with the CLAP for head-on collisions forming one of these wormholes.  The total radiated energy is slightly off the
predictions of the fit \eqref{eq:rad_fit} because points $\mathfrak{R}\gtrsim 0.8$ were not used in the fit.  Notice
that the total radiated energy is over one order of magnitude larger than the corresponding process forming a BH: 
substantial amount of energy is released in late-time echoes.

Furthermore, a spectral analysis shows that the ECO formation excites certain characteristic frequencies, which
correspond to the quasinormal frequencies of the final object.  In the particular case of a thin-shell wormhole
described above, most of the radiation is in fact contained in such modes: the energy spectrum shows clear peaks at such
quasinormal frequencies, which are in excellent agreement with the spectral findings of Ref.~\cite{Cardoso:2016rao}.

\section{BBH\lowercase{s} and scalar fields}\label{sec:CLAP_STT}
From the first observation of the Higgs boson by the ATLAS Collaboration\til\cite{Aad:2012tfa}, a growing recognition
has been given in studying the effects of scalar particles, both at a cosmological and an astrophysical
level\til\cite{Ikeda:2018nhb,Boskovic:2018lkj,Isi:2018pzk,Bernard:2019nkv,Sun:2019mqb,Berti:2019wnn,Cardoso:2020hca,Ikeda:2020xvt}.
Moreover, most of the modifications of GR which have been proposed so far can be reformulated in terms of couplings between gravity and extra fields, the simplest of which are scalar fields\,\cite{Berti:2015itd}. In this Section we study how the presence of scalar fields may affect a BBH collision, and whether they can leave
observable imprints during the GWs generation.

We shall consider gravity minimally coupled with a complex scalar field. Since we are interested in BH solutions, we do not include matter fields in the model. Thus, the action is:
\begin{equation}
S=\int
d^4x\sqrt{-g}\left(\frac{R}{16\pi}-\frac{1}{2}g^{\mu\nu}\partial_\mu\Phi^*\partial_\nu\Phi\right)\,.\label{eq:action_minimal}
\end{equation}
The field equations obtained from this action are Einstein's equations coupled with the Klein-Gordon equations:
\begin{align}
\label{eq:Einstein_eq}
R_{\mu\nu}-\frac{1}{2}g_{\mu\nu}R&=8\pi T_{\mu\nu}\,,\\
\label{eq:KG_eq}
\square\,\Phi&=0\,,
\end{align}
where 
\begin{equation}
  T_{\mu\nu}=-\frac{1}{2}g_{\mu\nu}\left(\partial_\lambda\Phi^*\partial^\lambda \Phi\right)+
  \frac{1}{2}\left(\partial_\mu\Phi^*\partial_\nu\Phi+\partial_\mu\Phi\partial_\nu\Phi^*\right)\,,\nonumber
\end{equation}
is the scalar field stress-energy tensor. A wide class of modified gravity theories in which gravity is non-minimally coupled with a scalar field - the so-called Bergmann-Wagoner scalar-tensor theories (see e.g.~\cite{fujii2003scalar,Berti:2015itd} and references therein), is formally equivalent to the
theory in Eq.~\eqref{eq:action_minimal}\footnote{The theories are related through a conformal rescaling of the metric~\cite{fujii2003scalar}.}, if restricted to vacuum spacetimes. Thus, the scalar field $\Phi$ can be interpreted either as a fundamental ``matter'' field in GR, or as a gravitational degree of freedom in a
modified gravity theory.

The $3+1$ decomposition of the action~\eqref{eq:action_minimal} has been discussed in Sec.~\ref{sec:3+1}.
In particular, the constraint equations have the form
\begin{align}
\label{eq:Hstt1}
&\mathcal{H}^{\rm GR}-16\pi \rho=0\,,\\
\label{eq:Mstt1}
&\mathcal{M}^{\rm GR}_i-8\pi j_i=0\,,
\end{align}
where the energy density $\rho$ and the energy-momentum flux $j_i$ of the scalar field are
given in Eqs.~\eqref{eq:rhostt}-\eqref{eq:jstt}.

The theories described by the action~\eqref{eq:action_minimal} satisfy the {\it no-hair theorem}: stationary BH
solutions are described by the Kerr metric, and thus they have vanishing scalar
field\,\cite{Bekenstein:1972ky,Hawking:1972qk,Sotiriou:2011dz} (see also~\cite{Cardoso:2016ryw} and references
therein). Therefore, we know that the remnant of a BBH collision becomes --~in the timescale of the QNM oscillations,
i.e. of $\sim1-10\,M$~-- a stationary BH solution, with vanishing scalar field.

The no-hair theorem does not tell us what happens before reaching the final
stationary configuration: it does not constrain the {\it dynamics} of BH spacetimes. Still, a similar result applies to
the inspiral part of a BBH coalescence: an analysis in the post-Newtonian (PN)
approximation~\cite{Blanchet:2013haa,Poisson_will_2014}, which accurately describes the BBH inspiral, shows that the
binary dynamics in the theory~\eqref{eq:action_minimal}, up to $2.5$ order in the PN expansion, is the same as in
GR~\cite{Will:1989sk}\,\footnote{As argued before, similar results hold also in the PN treatment of Bergmann-Wagoner
  scalar-tensor gravity.}.  However, we do not know whether the presence of a scalar field significantly affects the BBH
dynamics during the merger and ringdown stages.

In order to address this problem, we shall study the QNMs of the scalar field during the merger and ringdown of a BBH (head-on) collision. We do not expect the scalar field to grow large before the collision, thus we shall treat it as a
perturbation of the BBH spacetime; we define a perturbation parameter $\epsilon\ll1$, such that $\Phi=O(\epsilon)$.
Therefore, $T_{\mu\nu}=O(\epsilon^2)$ and, to linear order in the perturbation, we can neglect the scalar field from
Einstein's equations~\eqref{eq:Einstein_eq} (and in particular from the constraint equations~\eqref{eq:Hstt1}-\eqref{eq:Mstt1}). We shall then study the linearized ($O(\epsilon)$) field equation of the scalar field in the BBH spacetime, which is modeled using the CLAP approximation. For such configurations we shall compute the scalar field QNMs. Then, by
comparing the scalar field QNMs in a BBH spacetime with the scalar field QNMs in a stationary BH spacetime, we will
assess whether the binary dynamics significantly affects the scalar field dynamics.

We describe the BBH spacetime (neglecting the scalar field, as discussed above) using BL initial data. Therefore, we
recast the BBH spacetime as a perturbation of the Schwarzschild metric. Including the leading-order quadrupolar
contribution, as discussed in Sec.~\ref{subsec:recast_BH}, the perturbed spacetime can be written as
$g_{\mu\nu}=g^{(0)}_{\mu\nu}+h_{\mu\nu}$ where $g^{(0)}_{\mu\nu}={\rm diag}(-f,f^{-1},r^2,r^2\sin^2\theta)$ and
$h_{\mu\nu}$ is given by Eqs.~\eqref{eq:BLrecast}-\eqref{eq:H2_Kl_equality}:
\begin{align}
h_{rr}&=f^{-1}gP_2(\cos\theta)\xi_2\,,\nonumber\\
h_{\theta\theta}&=r^2g P_\ell(\cos\theta)\xi_2\,,\label{eq:BLrecast1}
\end{align}
where $g=4\left(1+M/(2R)\right)^{-1}M^3/R^3$, the isotropic coordinate $R$ is defined in
Eq.\til\eqref{eq:R_isotropic_to_Schw}, and $\xi_2=Z_0^2/(2M^2)\lesssim1$.
We stress that we are assuming the background to be stationary, thus neglecting the motion of the BHs, in the timescale of the
oscillation ($t\sim(1-10)M$). This is a crude approximation, since the BH separation changes, and their
velocities become non-negligible in this timescale. Thus, the result of this computation should be considered as an
order-of-magnitude estimate of the effect of the BH dynamics on the scalar QNMs.

Since $h_{\mu\nu}\propto Z_0^2$, we can expand the D'Alembertian operator $\square=\frac{1}{\sqrt{-g}}\partial_\mu\left(
g^{\mu\nu}\sqrt{-g}\partial_\nu\right)$ in Eq.~\eqref{eq:KG_eq}, for small separations, as
\begin{equation} \label{eq:box_expansion}
\square = \square^{(0)}+ Z_0^2 \, \square^{(1)} + \mathcal{O}(Z_0^3)\,,
\end{equation}
where the explicit form of the operator $\square^{(1)}$ is given in Appendix~\ref{app:separateKG}.
Expanding the scalar field as
\begin{equation}\label{eq:spher_harm_decomp}
\Phi\left( t,r,\theta,\phi\right)=\frac{1}{r}\sum_{\ell,m}\psi_{\ell m}\left( t,r\right) Y^{\ell m}\left(\theta,\phi\right)\,,
\end{equation}
we get a non-separable equation, where the harmonic component $\psi_{\ell m}$ couples with the components
$\psi_{\ell\pm2\,m}$. As discussed in Appendix~\ref{app:separateKG}, we can follow the same approach used to study
scalar field perturbations around rotating BHs (see e.g.~\cite{Cano:2020cao,Pierini:2021jxd}); indeed, the leading-order
rotational corrections are quadrupolar as well. Remarkably, the $\ell\leftrightarrow\ell\pm2$ couplings do not affect
the QNM frequencies at leading order in the perturbations (see also~\cite{kojima1993normal,Pani:2012bp}), and thus they
can be neglected, leading to a decoupled, Schr\"{o}dinger-like equation:
\begin{align} \label{eq:KG_notortoise}
  &\frac{\partial^2 \psi_{lm}}{\partial t^2}+\frac{\partial^2 \psi_{lm}}{\partial r^2}
  \left( U_0+Z_0^2\tilde{U}_0\right)+\frac{\partial \psi_{lm}}{\partial r}  \left( U_1+Z_0^2\tilde{U}_1\right)\nn\\
&+\psi_{lm} \left(  W_0 +Z_0^2  W_1\right)=0\,,
\end{align}
where $\ell\ge1$\,\footnote{The monopolar $\ell=0$ perturbations are not affected by the $Z_0^2$ corrections (see
  Eq.\til\eqref{eq:q12lm}), hence the $\ell=0$ QNMs are the same as in the single BH case.}.  The derivation of
Eq.~\eqref{eq:KG_notortoise} and the explicit form of the functions $U_A(r)$, $\tilde U_A(r)$, $W_A(r)$ ($A=0,1$) are given in
Appendix~\ref{app:separateKG}.

In order to find the QNMs, we have solved Eq.~\eqref{eq:KG_notortoise} through direct integration with outgoing boundary
conditions at infinity and with ingoing boundary conditions at the horizon. The boundary conditions have been determined as a perturbative, polynomial expansion at each boundary, whose coefficients have been found solving
Eq.~\eqref{eq:KG_notortoise} order by order, as explained, e.g. in Ref.\til\cite{Pani:2013pma}.

\begin{figure}
\centering
\includegraphics[width=8.5cm,keepaspectratio]{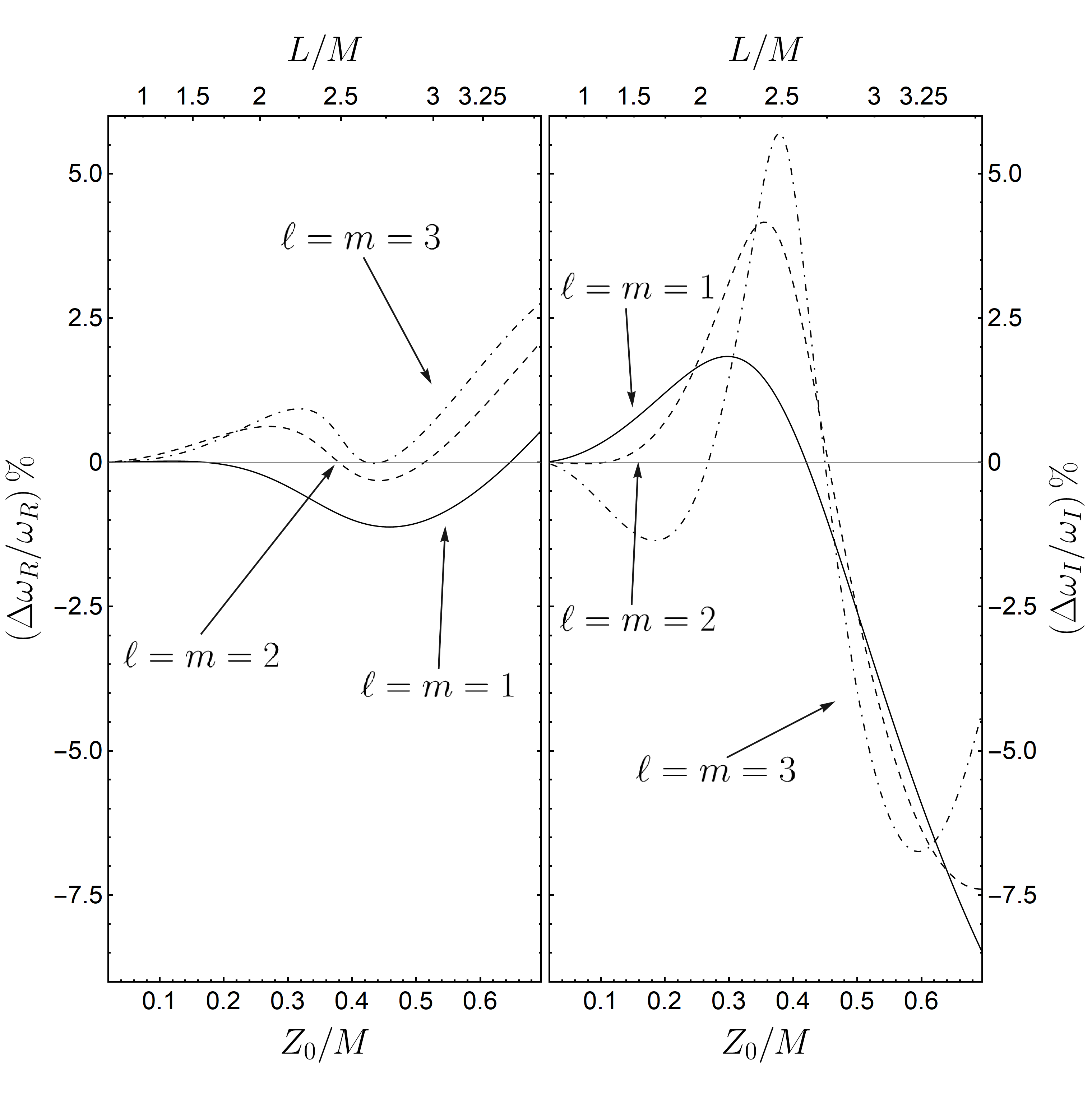}
\caption{Fractional percentage shifts $\Delta\omega_R/\omega_R$ (left panel) and $\Delta\omega_I/\omega_I$ (right
  panel), as defined in Eq.~\eqref{eq:shifts}, for $\ell=1,2,3$ and $m=\ell$, as functions of the BHs initial separation
  $Z_0/M$, or, equivalently, $L/M$ according to Eq.\til\eqref{eq:defL}.}
\label{fig:BBH_QNM}
\end{figure}

To validate our integration method, as a first step we computed the fundamental scalar QNMs of a Scwharzschild BH,
setting $Z_0=0$. Our results agree with those in the literature (e.g.~\cite{GRITJHU}) within $0.2\%$. Then, we
computed the scalar QNMs in BBH collisions for different values of $\ell\ge1$ and of $Z_0$. Figure~\ref{fig:BBH_QNM} shows the fractional percentage shift of real and imaginary parts of the QNMs with respect to those in Schwarzschild BHs:
\begin{equation}
\frac{\Delta\omega_{R/I}}{\omega_{R/I}}=\frac{\omega_{R/I}-\omega_{R/I}^{\rm (Schw)}}{\omega_{R/I}^{\rm (Schw)}}\,,\label{eq:shifts}
\end{equation}
for $\ell=1,2,3$ and $m=\ell$, as functions of $Z_0/M\le0.7$.

As we can see from Fig.~\ref{fig:BBH_QNM}, the QNMs shifts have a non-trivial dependence on the BH separation $Z_0$, but
each mode is shifted by just a few percent from the corresponding mode of an isolated BH. Despite the approximations
discussed above, this result provides a strong indication that the BH dynamics does not significantly affect the
behaviour of the scalar perturbations, at linear order. In particular, our results suggest that the
scalar modes do not become unstable during the BH coalescence, and thus a small scalar field would remain small during
the merger.
For results concerning non-trivial effects of a self-interacting ambient scalar field on a BBH, we refer the reader to Refs.~\cite{Healy:2011ef,Berti:2013gfa}.

Thus, our results indicate that scalar fields in GR, or in modified gravity theories in which the no-hair theorem
applies, do not significantly affect the coalescence of BBHs.

These results are also interesting in light of a completely different question: at what time the GW signal from a BBH
coalescence can be described as a superposition of QNMs? This problem, i.e. the determination of the starting time of
the ringdown, is widely
debated~\cite{Berti:2007fi,Baibhav:2017jhs,Bhagwat:2017tkm,Ota:2019bzl,Giesler:2019uxc,Forteza:2020hbw}. Indeed, the
procedure of constructing a GW template is based on joining different approximations, from the regime where the BHs
separation is large (PN approximation) to the ringdown oscillations (perturbation theory), passing through the highly
non-linear merger process that requires numerical approaches. Our results indicate that an observer measuring the scalar
oscillations during a BBH collision may extend the validity of the ringdown treatment closer to the merger, where, in
principle, only full numerical studies are accurate and reliable. The direct detection of scalar fields oscillations is not possible with current detectors, if one assumes GR as the correct theory of gravitation. However, we may take this as an indication that the same may hold true for gravitational waveforms, for which, instead, detections are possible through current GW interferometers.

\section{Conclusions and outlook}\label{sec:Conclusions}
GW astronomy has the potential to answer crucial questions regarding the correct description of gravity. The full
exploitation of such potential requires knowledge about the dynamics of compact objects in a generic theory of gravity.
While NR is the tool of excellence for this, the evolution of a single binary within the context of a modified theory
can take months to perform on supercomputers, and may require years of careful study of the relevant partial
differential equations and associated well-posedness.

In this work, we explored the close-limit approximation as a ``quick-and-dirty'' tool to understand nonlinear
coalescence processes. Its remarkable agreement with full nonlinear simulations is an important benchmark. In fact,
albeit it is a perturbative scheme, it uses constraint-satisfying initial data, and their evolution works accurately
even when the premises of the model are only partially satisfied. Consequently, this provides some confidence that this
technique works well also when extending those studies beyond GR or BH spacetimes. The main requirement to use the CLAP
consists in having solutions of the constraint equations. These can be solved, as we showed, also in the presence of
fundamental fields (see also Ref.~\cite{Okawa:2014nda,Zilhao:2015tya}).

This work is just a first step towards the implementation of the
CLAP beyond the standard GR/BH scenario. Our computation is based on strong approximations; thus our results should be
interpreted as qualitative indications.
Further effort is required here. We expect the payoff to be significant:
with much less computational time and effort one is able to investigate setups that describe better nonlinear
geometries. We showed how the CLAP can work for the coalescence of compact, horizonless objects, and how it too predicts
the existence of echoes in gravitational waves. This is a significant result in that it extends and complements other
past perturbative calculations~\cite{Cardoso:2016rao,Cardoso:2016oxy,Cardoso:2019rvt}.
Moreover, we studied scalar fields minimally coupled with gravity in BH spacetimes (which are
equivalent to non-minimally coupled scalar fields in a Bergmann-Wagoner scalar-tensor theory), estimating the scalar
modes in the merger of a BH binary, and showing that they are very similar to those in a stationary BH
spacetime. 

In future works, we will relax some of the approximation used in this work, by considering
non-head-on collisions of spinning objects, and non-stationary initial data. Moreover, we shall use the CLAP to
investigate compact objects collisions in other - and perhaps more complicated, modified gravity theories such as
Einstein-scalar-Gauss-Bonnet gravity.

\noindent{\bf{\em Acknowledgements.}}
%
L.~A. acknowledges financial support provided by Funda\c{c}ao para a Ci\^{e}ncia e a Tecnologia Grant number
PD/BD/128232/2016 awarded in the framework of the Doctoral Programme IDPASC-Portugal.
V.~C.\ acknowledges financial support provided under the European Union's H2020 ERC 
Consolidator Grant ``Matter and strong-field gravity: New frontiers in Einstein's 
theory'' grant agreement no. MaGRaTh--646597.
This project has received funding from the European Union's Horizon 2020 research and innovation programme under the Marie Sklodowska-Curie grant agreement No 101007855.
We thank FCT for financial support through Project~No.~UIDB/00099/2020.
We acknowledge financial support provided by FCT/Portugal through grants PTDC/MAT-APL/30043/2017 and PTDC/FIS-AST/7002/2020.
We also acknowledge support from the Amaldi Research Center funded by the MIUR program "Dipartimento di Eccellenza"
(CUP: B81I18001170001), and from the MIUR grant PRIN2017-MB8AEZ.
The authors would like to acknowledge networking support by the GWverse COST Action 
CA16104, ``Black holes, gravitational waves and fundamental physics.''
%

\appendix

\section{Numerical integration of the Zerilli equation with source} \label{app:Zerilli_solver}
We shall discuss the numerical integration of the Zerilli equation with source in the frequency domain, Eq.~\eqref{eq:zerS}
(see also~\cite{Lousto:1996sx,Campanelli:1997un}):
\begin{equation} \label{eq:Zerilli_model}
\frac{\partial^2 \tilde{\psi}}{\partial r_*^2}+ \left(\omega^2-V_Z\right)\tilde{\psi}=S\,,
\end{equation}
where $\tilde\psi(\omega,r)$ is the Laplace transform~\eqref{eq:laplace_psi} of the Zerilli function
$\psi(t,r)$,
\begin{equation}
  \tilde\psi(\omega,r_*)=\int_{0}^\infty dt\psi(t,r_*)e^{i\omega t}\,,\label{eq:laplace_psi2}
\end{equation}
and
\begin{equation}
S(\omega,r)=i\omega\psi(t=0,r)=i\omega\sqrt{\frac{4\pi}{5}}\frac{1}{1+\frac{3M}{2r}}Q(r)\xi_2\,,\label{eq:sourceterm}
\end{equation}
where $Q(r)$ is given in Eq.~\eqref{eq:defQ} and $\xi_2=Z_0^2/(2M^2)\lesssim1$.

We shall find the solution of Eq.~\eqref{eq:Zerilli_model} satisfying ingoing boundary conditions at the horizon,
outgoing boundary conditions at infinity, by employing two different approaches: the Green function approach and a
shooting method, finding the same results.

\subsection{\em Collision to a Schwarzschild BH}\label{app:solveZerBH}
If the Zerilli equation~\eqref{eq:Zerilli_model} describes perturbations of a Schwarzschild BH, it is defined in
$-\infty<r_*<+\infty$. In this case, the source term~\eqref{eq:sourceterm} does not vanish at the horizon,
$S(r_*\to-\infty)=i\omega\psi(t=0,r_*\to\infty)=\bar S\neq0$. The ingoing wave boundary conditions at the horizon
$r_*\to-\infty$, $\partial\psi/\partial r_*=\partial\psi/\partial t$\,\footnote{Note that these conditions are consistent with $\dot\psi(t=0)=0$
  beacuse in the BL initial data $\partial\psi/\partial r_*=0$ at the horizon.} translate in the Laplace transform space into
$\partial\tilde{\psi}/\partial r_*(\omega)=-i\omega\tilde{\psi}(\omega)-\psi(t=0)$ . Therefore, the boundary conditions of
$\tilde\psi(\omega,r_*)$ are:
\begin{align}
\tilde\psi(\omega,r_*)&=A^He^{-i\omega r_*}+\frac{\bar S}{\omega^2}~~~~~(r_*\to-\infty)\,,\nonumber\\
\tilde\psi(\omega,r_*)&=A^\infty e^{i\omega r_*}~~~~~~~(r_*\to+\infty)\,,
  \label{eq:bcsomm1}
\end{align}
with $A^H$, $A^\infty$ constants to be determined. The constant term is related to the fact that the BL initial data do
not vanish at the horizon, and do not affect the GW emission at infinity.

The Green function approach consists in finding two independent solutions of the homogeneous Zerilli equations:
$\tilde{\psi}^H$, satisfying ingoing wave conditions at the horizon, and $\tilde{\psi}^\infty$, satisfying outgoing wave
conditions at infinity, i.e.
\begin{align}
& \hspace{-0,5cm}\tilde{\psi}^H= \begin{cases}
       & e^{-i \omega r_*}, \qquad \qquad \qquad \qquad r_*\rightarrow -\infty\,,\\
       & D_{\rm in}e^{-i \omega r_*}+D_{\rm out} e^{+i \omega r_*},\,\,\,\, r_*\rightarrow +\infty\,,
    \end{cases} \nn\\
  & \label{eq:homogeneous_sols_BBH}
  \hspace{-0,5cm}\tilde{\psi}^\infty= \begin{cases}
       & B_{\rm in}e^{-i \omega r_*}+B_{\rm out} e^{+i \omega r_*},\,\,\, r_*\rightarrow -\infty\,,\\
       & e^{i \omega r_*},  \qquad \qquad \qquad \,\,\, \qquad r_*\rightarrow +\infty\,.
    \end{cases} 
\end{align}
The solution of the Zerilli equation with source, Eq.~\eqref{eq:Zerilli_model}, satifying the boundary
conditions~\eqref{eq:bcsomm1}, is then:
\begin{equation} \label{eq:Zerilli_gensol}
  \tilde{\psi}(\omega,r_*)=\frac{\tilde{\psi}^\infty}{W} \int_{-\infty}^{r_*}  S
  \tilde{\psi}^H dr'_*+\frac{\tilde{\psi}^H}{W} \int_{r_*}^\infty  S \tilde{\psi}^\infty dr'_*\,,
\end{equation}
where $W=\tilde{\psi}^H\left( \partial\tilde{\psi}^{\infty}/\partial r_*\right)-\tilde{\psi}^\infty\left( \partial
\tilde{\psi}^{H}/\partial r_*\right)$
is the constant Wronskian of the homogeneous equation. By imposing the boundary conditions at infinity~\eqref{eq:bcsomm1}
we find
\begin{equation}
  A^\infty(\omega)=\frac{1}{W}\int_{-\infty}^{+\infty}S\tilde{\psi}^Hdr_*\label{eq:zerinf}\,.
\end{equation}

Apparently, the integral~\eqref{eq:zerinf} is not well defined at the lower bound, where the integrand reduces to the
oscillating term $\bar Se^{-i\omega r_*}$. This is due to the fact that, strictly speaking, the Laplace
transform~\eqref{eq:laplace_psi2} is well defined in the upper complex plane; the inverse Laplace transform can be
computed along a path $\omega=\omega_R+i\epsilon$ with $\epsilon\ll1$ and $-\infty<\omega_R<+\infty$. Thus, along this
path $e^{-i\omega r_*}\to0$ as $r_*\to-\infty$ and the oscillating term disappears. As suggested
in~\cite{Lousto:1996sx}, we can compute the integrals for real values of $\omega$, as long as we subtract the ill-valued
contribution at the horizon:
\begin{equation}
  A^\infty(\omega)=\frac{1}{W}\int_{\bar{r}_*}^{+\infty}S\tilde{\psi}^Hdr_*+\frac{i}{\omega}\frac{\bar{S}}{W}e^{-i\omega \bar{r}_*}\,,
  \label{eq:zerinf_ren}
\end{equation} 
where $\bar{r}_*$ is negative and very large. We have computed $A^\infty(\omega)$, by evaluating the integrals from
$\bar{r}_*=-44\,M$ to the extraction radius $r_*^{\rm extr}=400\,M$.

The time-domain Zerilli function at infinity can then be computed, as a function of the retarded time $u=t-r_*$, as
\begin{equation}
\psi(u)=\frac{1}{2\pi}\int_{-\infty}^{+\infty}A^\infty(\omega) e^{-i\omega u}d\omega\,.\label{eq:invlapl}
\end{equation}

The shooting approach, instead, consists in the numerical integration of Eq.~\eqref{eq:Zerilli_model}, from
$\bar{r}_*=-44\,M$ to the extraction radius $r_*^{\rm extr}=400\,M$, by imposing the boundary conditions~\eqref{eq:bcsomm1}
and matching the solution at $r_*^{\rm extr}$ with an analytic expression obtained by an asymptotic expansion of
Eq.~\eqref{eq:Zerilli_model}.
For each value of the frequency $\omega$, we performed the numerical integration of Eq.~\eqref{eq:Zerilli_model} for
different values of $A^H$ until we obtained an outgoing wave at infinity, as in Eq.~\eqref{eq:bcsomm1}. In this
way we computed the function $A^\infty(\omega)$ and then, by Eq.~\eqref{eq:invlapl}, the Zerilli function at infinity.

The results of the two approaches perfectly agree with each other, and they also agree with the results of~\cite{Price:1994pm}.

\subsection{\em Collision to an Extreme Compact Object}\label{app:solveZerECO}
If the outcome of the collision is an ECO, the Zerilli equation~\eqref{eq:Zerilli_model} describes perturbations of the
ECO spacetime which, as discussed in Sec.~\ref{sec:BH_mimickers}, coincides with Schwarzschild's spacetime with the
domain restricted to $r_{0*}<r_*<+\infty$. Moreover, we impose at $r_*\to r_{0*}$ the partially reflecting boundary
conditions ~\eqref{eq:BC_ECO}:
\begin{align}
  \tilde{\psi}&=A^H(e^{-i \omega(r_*-r_{0*})}+\mathfrak{R}e^{i \omega(r_*-r_{0*})})~~~(r_*\to r_{0*})\nonumber\\
&+\frac{\hat S}{\omega^2} \label{eq:BC_ECO2}\\
  \tilde{\psi}&=A^\infty e^{i \omega r_*}~~\hspace{3.7cm}(r_*\to\infty)  \nonumber
\end{align}
with $\mathfrak{R}$ reflectivity coefficient which we assume, for simplicity, to be constant, and the constant term in
the ingoing boundary conditions is due to the fact that $S(r_{0*})=i\omega\psi(t=0,r_{0*})=\hat S\neq0$.  Note that
since the tortoise coordinate does not extend to $-\infty$, the Laplace transform $\tilde\psi(\omega,r_*)$ is
well-defined for real frequencies, and we do not need to worry about ill-defined contributions to the integrals.

In this case we only perform the integration using the shooting approach, i.e. we integrate
Eq.~\eqref{eq:Zerilli_model}, from $r_{0*}$ to the extraction radius $r_*^{\rm extr}=400\,M$, by imposing the
boundary conditions~\eqref{eq:BC_ECO2}, for different values of $A^H$, until we obtained an outgoing wave at
infinity. Then, using Eq.~\eqref{eq:invlapl} we compute the Zerilli function at infinity.

\section{Perturbative expansion of the Klein-Gordon equation}\label{app:separateKG}
Let us consider the Klein-Gordon equation for a massless scalar field,
\begin{equation}
\square\Phi=\frac{1}{\sqrt{-g}}\partial_\mu\left(g^{\mu\nu}\sqrt{-g}\partial_\nu\right)\Phi=0
\end{equation}
on the binary BH spacetime described by the BL solution linearized in the quadrupole contribution,
$g_{\mu\nu}=g^{(0)}_{\mu\nu}+h_{\mu\nu}$ with $g^{(0)}_{\mu\nu}$ Schwarzschild metric and $h_{\mu\nu}$ given in
Eq.~\eqref{eq:BLrecast1}. Expanding the D'Alembertian operator in powers of the parameter $Z_0$ as in
Eq.~\eqref{eq:box_expansion}, neglecting the terms $O(Z_0^3)$, it has the form:
\begin{equation} \label{eq:KG_pert}
\left( \square^{(0)}+ Z_0^2 \, \square^{(1)} \right) \Phi =0\,.
\end{equation}
We expand the scalar field in scalar spherical harmonics $Y^{\ell m}(\theta,\phi)$, as in
Eq.~\eqref{eq:spher_harm_decomp}. Since Schwarzschild's spacetime is spherically symmetric, the spherical harmonics are
eigenfuctions of the the D'Alembertian on $g^{(0)}_{\mu\nu}$, $\square^{(0)}$, while the operator $\square^{(1)}$
couples harmonics with different values of $\ell$ (but do not couple harmonics with different values of $m$).

We shall look for a solution which, in the limit $Z_0\to0$, has given harmonic indexes $(\ell,m)$. The contributions
with $\ell'\neq\ell$ are of order $O(Z_0^2)$, and we can write
\begin{align} \label{eq:Phi_ansatz_2}
\Phi&=\frac{\psi_{\ell\,m}\left( t,r\right) Y^{\ell\,m}\left(\theta,\phi\right)}{r}\nn\\
&+Z_0^2 \sum_{\ell'\neq\ell}\frac{\psi_{\ell'\, m}\left( t,r\right) Y^{\ell'\, m}\left(\theta,\phi\right)}{r}\,.
\end{align}
Replacing Eq.\til\eqref{eq:Phi_ansatz_2} in Eq.\til\eqref{eq:KG_pert} we find 
\begin{align} \label{eq:KG_pert_2}
& \square^{(0)} \left[\frac{\psi_{\ell\, m} Y^{\ell\,m} \left(\theta,\phi\right)}{r}\right]+ Z_0^2  \square^{(1)} \left[ \frac{\psi_{\ell\,m} Y^{\ell\,m} \left(\theta,\phi\right)}{r} \right]\nn\\
&+ Z_0^2 \sum_{\ell'\neq \ell} \square^{(0)} \left[ \frac{\psi_{\ell'\,m} Y^{\ell'\,m} \left(\theta,\phi\right)}{r}\right] =0\,.
\end{align}
Since the spherical harmonics are eigenfunctions of the operator $\square^{(0)}$, by projecting the above equation on
the complete basis of spherical harmonics, the components $\psi_{\ell'\,m}$ with $\ell'\neq\ell$ vanish, and we obtain a
{\it decoupled equation}, whose solution, with appropriate boundary conditions, gives the ``deformed'' QNMs
corresponding to the harmonic indexes $\ell,m$. We note that this result is analogous to the case of slowly rotating
spacetimes, whose QNMs - at leading order in the rotation - are not affected by the couplings between different harmonic
components~\cite{kojima1993normal,Pani:2012bp}. The $O(Z_0^2)$ term of Eq.~\eqref{eq:KG_pert_2} can be written as
\begin{align}
 &  \square^{(1)} \left[ \frac{\psi_{\ell\,m} Y^{\ell\,m} \left(\theta,\phi\right)}{r} \right]=-\frac{\partial Y^{\ell\, m}}{\partial \theta}\frac{3 g \sin (2 \theta ) }{8 M^2 r^3}\psi_{\ell\,m}\nn\\
  & +Y^{\ell\, m}\Bigg(- r\left(-r^2 f dg/dr+4 M g\right)\frac{\partial \psi_{\ell\,m}}{\partial r}-2r^3 f g\frac{\partial^2 \psi_{\ell\,m}}{\partial r^2}\nn\\
&+\left(-r^2 f dg/dr+2 g (\ell (\ell+1) r+2 M)\right)\psi_{\ell\,m}\Bigg)\frac{\alpha\left(\theta\right)\,}{16 M^2 r^4} \,,
\label{eq:KGZ02}
\end{align}
where $\alpha\left(\theta\right)=1+3 \cos (2 \theta )$ and we remind that $f=1-2M/r$. The projection on $Y^{\ell\,m}$ of the $O(0)$ terms in Eq.~\eqref{eq:KG_pert_2} yields the standard scalar field equation in
Schwarzschild spacetime, 
\begin{align} 
  &-\frac{1}{r f}\Bigg(\frac{\partial^2\psi_{\ell\,m}}{\partial t^2}-f^2\frac{\partial^2\psi_{\ell\, m}}{\partial r^2}
  -f\frac{df }{dr}\frac{\partial\psi_{\ell\, m}}{\partial r}\nn\\
&+f\frac{\ell (\ell+1) r+2 M }{r^3}\psi_{\ell\, m}\Bigg)\,.
\end{align}
If we include the projection on $Y^{\ell\,m}$ of the $O(Z_0^2)$ terms, i.e. of
Eq.~\eqref{eq:KGZ02}, Eq.~\eqref{eq:KG_pert_2} gives
\begin{align} \label{eq:KG_notortoise_app}
  &\frac{\partial^2 \psi_{\ell\,m}}{\partial t^2}+\frac{\partial^2 \psi_{\ell\,m}}{\partial r^2}
  \left( U_0+Z_0^2\tilde{U}_0\right)+\frac{\partial \psi_{\ell\,m}}{\partial r}  \left( U_1+Z_0^2\tilde{U}_1\right)\nn\\
&+\psi_{\ell\,m} \left(  W_0 +Z_0^2  W_1\right)=0\,,\nn\\
&
\end{align}
where 
\begin{align}
\label{eq:potentials_notortoise}
&U_0(r)=-f^2\,,\nn\\
&\tilde{U}_0(r)=\frac{\left( r-2 M \right) f q^{(1)}_{\ell\,m} g}{8M^2r}\,,\nn\\
&U_1(r)= -f \frac{df}{dr} \,,\nn\\
& \tilde{U}_1(r)= -\frac{q^{(1)}_{\ell\,m} (2 M-r) \left(4 M g(r)-f r^2 dg/dr\right)}{16 M^2 r^3}\,,\nn\\
&W_0(r)=f\frac{\ell (\ell+1) r+2 M }{r^3}\,,\nn\\
&W_1(r)=\frac{q^{(1)}_{\ell\,m}}{16 M^2 r^4} (f r^2 (r-2 M) dg/dr\nn\\
&+2 g (2 M-r) (l (l+1) r+2 M))+3q^{(2)}_{\ell\,m}\frac{(r-2M)g}{4M^2r^3}\,,
\end{align}
and we have defined the constant coefficients
\begin{align} \label{eq:q1lm}
q^{(1)}_{\ell\,m}  \equiv&\int d\Omega \left( Y^{\ell\,m}\right)^* Y^{\ell\,m} \alpha
  \,,\\ \label{eq:q2lm}
  q^{(2)}_{\ell\,m}\equiv &\int d\Omega \sin \theta \cos \theta \left( Y^{\ell\,m}\right)^*
  \frac{d Y^{\ell\,m}}{d\theta}  \,.
\end{align}
Note that
\begin{equation}
\label{eq:q12lm}
q^{(1)}_{00}=q^{(2)}_{00}=0\,,
\end{equation}
therefore the $\ell=0$ equation is not affected by the $O(Z_0^2)$ corrections. For $0<\ell\le2$, the quantities
$q^{(1,2)}_{\ell\,m}$ are:
\begin{align} 
&  q^{(1)}_{1-1}=q^{(1)}_{11}=-\frac{4}{5},\,q^{(1)}_{10}=\frac{8}{5}\,,\nonumber\\
&  q^{(1)}_{2-2}=q^{(1)}_{22}=-\frac{8}{7},\,q^{(1)}_{2-1}=q^{(1)}_{21}=\frac{4}{7},\,q^{(1)}_{20}=\frac{8}{7}\,,\nonumber\\
& q^{(2)}_{1-1}=q^{(2)}_{11}=\frac{1}{5},\,q^{(2)}_{10}=-\frac{2}{5}\,,\nonumber\\
& q^{(2)}_{2-2}=q^{(2)}_{22}=\frac{2}{7},\,q^{(2)}_{2-1}=q^{(2)}_{21}=-\frac{1}{7},\,q^{(2)}_{20}=-\frac{2}{7}\,.
\end{align}

\newpage

\bibliographystyle{apsrev4}

\bibliography{References}

\end{document}